\documentclass[%
 aip,
 jcp,%
 amsmath,amssymb,
 reprint,%
]{revtex4-1}
\usepackage{graphicx}
\usepackage{amsmath}
\usepackage{amssymb}
\usepackage{dcolumn}% Align table columns on decimal point
\usepackage{bm}% bold math

\newcommand{\dd}{\textrm{d}}

\usepackage{color}

\begin{document}

\title{Quantum Monte Carlo with variable spins: fixed-phase and fixed-node approximations}

\author{Cody A. Melton and Lubos Mitas}
%\affiliation{
%Center for High Performance Simulation and Department of Physics
%North Carolina State University, Raleigh, NC 27695}
\affiliation{
Department of Physics, North Carolina State University, Raleigh, North Carolina 27695-8202, USA\\
}

\date{\today}

\begin{abstract}
We study several aspects of the recently
introduced  fixed-phase spin-orbit diffusion Monte Carlo (FPSODMC) method, in particular, its relation to the fixed-node method and its potential use as a general approach
for electronic structure calculations. 
We illustrate constructions of spinor-based 
 wave functions with the full space-spin symmetry without assigning up or down spin labels to particular electrons, effectively
  ``complexifying" even ordinary real-valued wave functions. Interestingly, with proper choice of the simulation parameters and spin variables, such fixed-phase calculations enable one to reach also the fixed-node limit. 
 The fixed-phase solution  provides a straightforward
 interpretation  as the lowest
 bosonic state in a given effective potential generated
 by the many-body approximate phase. In addition, 
 the divergences present at real wave function nodes are smoothed out to lower dimensionality,
 decreasing thus the variation of sampled quantities and making the sampling also more straightforward. We illustrate some of these 
 properties on calculations of selected 
 first-row systems that recover the fixed-node results with quantitatively similar levels 
 of the corresponding biases. 
At the same time, the fixed-phase approach opens new possibilities
for more general trial wave functions with further
opportunities for increasing 
accuracy in practical calculations. 
\end{abstract}

%\pacs{02.70.Ss, 71.15.Nc, 71.15.-m; 02.70.Ss, 71.15.-m, 31.15.V- }
\maketitle
\section{Introduction}

Recently, we  introduced a projector quantum Monte Carlo (QMC) method for calculating quantum systems with both spatial and spin degrees of freedom \cite{melton}. 
The approach is based on an overcomplete representation for spin variables such that the sampling is similar to the spatial variables. Given our choice of spin representation, the method involves the fixed-phase
\cite{ortiz} approximation, hence its acronym fixed-phase {\underbar s}pin-{\underbar o}rbit/{\underbar s}pin{\underbar o}r diffusion Monte Carlo (FPSODMC). This approach enabled us to carry out QMC calculations of atoms and molecules with spin-orbit 
interactions in the spinor formalism including cases where 
high accuracy was needed for both spin-orbit 
and electron correlation effects. In a subsequent work
we explored simple cases of fixed-phase vs. fixed-node
\cite{qmcrev}
approximations in order to compare the corresponding biases in these two related possibilities \cite{melton2, melton3}. We constructed simple
cases where both fixed-phase and fixed-node conditions 
were equivalent or very similar and we found comparable 
biases in the total energies using the two approximations.

In this work we explore this direction further by
investigating a clear unification and smooth transition between these approaches. It has been
known for some time that the fixed-node approximation is a special 
case of the fixed-phase approximation. Our method makes this
relationship explicit through the construction of 
trial wave functions that in a particular limit recover
the fixed-node trial function. We use this property for 
QMC fixed-phase calculations of several systems 
and we directly compare the fixed-phase biases to the corresponding fixed-node biases. In addition, we 
explicitly show how one can obtain the fixed-node result
as a limit of the fixed-phase calculation. 
We elaborate on the advantages and disadvantages of the fixed-phase approach as far as further QMC developments 
are concerned.

\section{Fixed-Phase Spinor Diffusion Monte Carlo}\label{section:FPDMC}

Let us briefly outline the key notions of the FPSODMC approach: fixed-phase approximation, continuous spin 
representation and the corresponding importance 
sampling approach.

\subsection{Fixed-Phase Approximation }\label{section:FPDMC}
For complex wave functions
we present a brief sketch of  
the fixed-phase method (FPDMC) \cite{ortiz} and its relation to the fixed-node flavor of DMC. 

 Let us consider the Born-Oppenheimer Hamiltonian $H=-(1/2)\nabla^2 + V(\mathbf{R})$, where $\nabla = (\nabla_1,\nabla_2, \ldots, \nabla_N)$ and $V$ denotes the electron-ion and electron-electron Coulomb interactions. 
 We denote spatial configurations as $\mathbf{R} = (\mathbf{r}_1, \mathbf{r}_2, \ldots, \mathbf{r}_N) \in \mathbb{R}^{dN}$, where $N$ is the number of particles and $d$ is dimensionality (here we assume $d=3$). 
 We assume a complex  wave function $\Psi(\mathbf{R},\tau) = \rho(\mathbf{R},\tau)e^{i \Phi(\mathbf{R},\tau)}$ and substitute it into the imaginary-time Schr\"{o}dinger equation.  For the amplitude $\rho(\mathbf{R},\tau)$ and phase $\Phi(\mathbf{R},\tau)$ we obtain
\begin{eqnarray}
    \label{rho-eqn} -\frac{\partial \rho(\mathbf{R},\tau)}{\partial \tau} &=& \left[ T_{kin} + V(\mathbf{R}) + \frac{1}{2} \left| \nabla \Phi(\mathbf{R},\tau)\right|^2\right] \rho(\mathbf{R},\tau) \nonumber \\ \\
    -\frac{\partial \Phi(\mathbf{R},\tau)}{\partial \tau} &=& \left[ T_{kin}-\frac{\nabla \rho(\mathbf{R},\tau) \cdot \nabla}{\rho(\mathbf{R},\tau)} \right] \Phi(\mathbf{R},\tau)
    \label{phase-eqn}
\end{eqnarray}
where $T_{kin}=-(1/2)\nabla^2$.
 The fixed-phase approximation is given by imposing 
 the phase to be equal to the phase of trial or variational wave function $\Psi_T({\bf R})=\rho_T({\bf R})e^{i\Phi_T({\bf R})}$ that is independent of $\tau$ 
 \begin{equation}
 \Phi(\mathbf{R},\tau) \overset{!}{=} \Phi_T(\mathbf{R}).
 \end{equation}
 so that the second equation is not considered any further.
 On the other hand, the stationary trial phase enables us to solve the equation for the non-negative amplitude $\rho$ and the corresponding energy eigenvalue. Clearly, both are now dependent on the trial phase through the additional potential $V_{ph}=1/2 | \nabla \Phi_T(\mathbf{R}) | ^2$.
 
 \subsection{Fixed-phase upper bound property}
The fixed-phase approximation is variational since the repulsive potential $V_{ph}$
can only raise the energy for an approximate phase \cite{ortiz}. 
 This is easy to see from the energy expectation with $\rho\exp(i\Phi_T)$ that must be an upper 
bound to the exact energy for
an arbitrary symmetric $\rho\geq 0$. 
The accuracy of this method clearly depends on the accuracy of the trial phase and 
the convergence towards the exact eigenvalue scales with the square of the difference between the exact 
and approximate trial function. 

\subsection{Fixed-phase as a special case of the fixed-node in general}
 The fixed-phase approximation is a generalization of the more familiar fixed-node approximation, what can be demonstrated in several ways.
 Let us present perhaps the simplest such construction
 \cite{melton2}, where
 we add a complex amplitude to a real-valued $\Psi_T(\mathbf{R})$ as follows. We denote the nodes of $\Psi_T$ as the set of configurations
\begin{equation}
    \Gamma =\left\lbrace \mathbf{R}\in\mathbb{R}^{dN} | \Psi_T(\mathbf{R})=0 \right\rbrace. 
\end{equation}
Now we add to $\Psi_T$ another function
(for simplicity, a nonnegative bosonic ground state of $H$) 
\begin{equation}
\tilde\Psi_T = \Psi_T +i\varepsilon \Psi_B   
\end{equation}
Taking the limit $\varepsilon \to 0$ leads to\cite{melton2}
\begin{equation}
V_{ph}(\mathbf{R})= V_{\infty} \delta (\mathbf{R}-\mathbf{R}_{\Gamma})   
\end{equation}
where $\mathbf{R}_{\Gamma} \in \Gamma$ and  $V_{\infty}$ diverges as $\propto 1/\varepsilon^2$, therefore $V_{ph}$  enforces any wave function to vanish at the node $\Gamma$, i.e., it is equivalent to the fixed-node boundary condition. 

\subsection{Spin Representation}\label{section:spinrep}

Let us denote one-particle spinors as
\begin{equation}
\label{eqn:spinor}
\chi ({\bf r},s)=\alpha\varphi^{\uparrow}({\bf r})\chi^{\uparrow}(s)
 +\beta\varphi^{\downarrow}({\bf r})\chi^{\downarrow}(s)
\end{equation}
where $s$ is the coordinate of the spin projection along
the $z-$axis.  
In its minimal representation
the spin variables have
discrete values  $s=\pm 1/2$ so that for $S_z$ eigenstates 
 $\chi^{\uparrow}(1/2)=\chi^{\downarrow}(-1/2)=1$,
$\chi^{\downarrow}(1/2)=\chi^{\uparrow}(-1/2)=0$. 
Clearly, the spin configuration space is non-compact
and imposes potentially large variations of important
quantities during the stochastic updates. Besides the fluctuations of various quantities of interest
(local energy, drifts, values of the wave function, etc.)
the method looses its efficiency in
the many-particle limit.

One possibility to address this obstacle is to make the spin configuration space compact and continuous, which allows for continuous evolution as well as importance sampling \cite{qmcrev, melton2}. We choose an {\em overcomplete} spin representation through the utilization of a 1D ring (i.e. a $U(1)$ representation) with the lowest pair of degenerate, orthogonal eigenstates as follows:
\begin{equation}
\begin{array}{l r} \langle s_j | \chi^\uparrow \rangle = e^{i s_j},\;\;\; & \langle s_j | \chi^\downarrow \rangle = e^{-i s_j}  \ \end{array}
\end{equation}
where the spin variable $s_j \in [0,2\pi)$.
Clearly, the paths in this space are continuous and 
resemble paths for spatial coordinates.
 
 \subsection{Importance sampling}
Rewriting the Schr\"odinger equation in an integral form 
with importance sampling by $\rho_T$ leads to the following equation for the mixed distribution $g=\rho\rho_T$
\begin{equation}
    g(\mathbf{R}',t+\tau) = \int \dd \mathbf{R} \; {\rho_T(\mathbf{R}')\over \rho_T(\mathbf{R})}
    G(\mathbf{R}' \leftarrow \mathbf{R},\tau)g(\mathbf{R},t)
\end{equation}
which is well-known
from the fixed-node QMC \cite{qmcrev,melton2}.  

Spin variables are sampled by introducing a spin ``kinetic'' energy with a corresponding energy offset such that for all $s_i$, $i \in \{ 1,2,...,N \}$ we write 
\begin{equation}
    T_i^s = -\frac{1}{2\mu_s}\left[ \frac{\partial^2}{\partial s_i^2} +1\right].
\end{equation}
where $\mu_s$ is an effective mass.
The full Hamiltonian then becomes $H' = H+\sum_{i=1}^N T_i^s$. 
Clearly, $T_i^s \psi(\mathbf{r}_i,s_i) = 0$ due to the introduced offset so that there is no energy contribution from the spin Laplacian. 
The inclusion of the spin kinetic energy leads to the following importance sampled Green's function 
\begin{equation}
    \label{eqn:spin_space_greens} \widetilde{G}( \mathbf{X}'\leftarrow\mathbf{X};\tau) 
\simeq T_{\mathbf{X'},\mathbf{X}}
e^{-\tau[ E_L(\mathbf{X}) + E_L(\mathbf{X}')-2E_T]/2} 
\end{equation}
with
\begin{eqnarray}
    T_{\mathbf{X}',\mathbf{X}} &\propto& \exp\left[ \frac{-\left| \mathbf{R'}-\mathbf{R} - \tau \mathbf{v}_D^\mathbf{R}(\mathbf{R}) \right|^2}{2 \tau}\right] \nonumber \\ 
    &&\times \exp \left[ \frac{-\left| \mathbf{S}'-\mathbf{S}-\tau_s\mathbf{v}_D^{\mathbf{S}}(\mathbf{S})\right|^2}{2\tau_s}\right]
\end{eqnarray}
where ${\bf X} = ({\bf r}_1, {\bf r}_2, ...,
{\bf r}_N,s_1,s_2,...,s_N) =({\bf R},{\bf S})$. Here we have introduced a spin time step $\tau_s = \tau/\mu_s$ as well as
${\bf v}_D^{\bf R} = \nabla_\mathbf{R} \ln \rho_T(\mathbf{X})$ and ${\bf v}_D^{\bf S} = \nabla_\mathbf{S} \ln \rho_T(\mathbf{X})$ which correspond to the spatial and spin drifts.  The local energy is given by $E_L=\textrm{Re}[(H'\Psi_T)/\Psi_T^{*}]$ \cite{qmcrev,melton}. 

Note that there are two possible limiting  cases
with regard to $\tau,\tau_s$, namely, $\tau_s >>\tau$
and vice versa. If $\tau_s$ is much larger than  the
spatial step the spin degrees of freedom are  evolving much faster so that effectively the spins are integrated 
out for each spatial step. That guarantees to
 provide the fixed-phase limit and it is expected that it will lead to the largest bias. 
 Indeed, this is what we have observed \cite{melton2}.
 
 The opposite limit corresponds to very slow spin evolution so that the spin configuration appears as an almost static external field for (relatively) much faster  spatial evolution. In the results section we come to this point again and we show that in this mode the simulations will enable us to recover the fixed-node solutions.

\section{Trial Wave Functions}\label{section:wf}

The FPSODMC trial functions are built from spinors $\chi(\mathbf{r},s)=\alpha\varphi^{\uparrow}({\bf r})\chi^{\uparrow}(s)
 +\beta\varphi^{\downarrow}({\bf r})\chi^{\downarrow}(s) $
where orbitals $\varphi^{\uparrow}, \varphi_{\downarrow}$ are calculated in spinor-based DFT, HF/DF 
or correlated methods.
The full configuration space for particles is $\mathbf{X} = \{ (\mathbf{r}_1,s_1), \ldots, (\mathbf{r}_N,s_N)\} \in \mathbb{R}^{3N} \times [0,2\pi)^{N}$ and
we write the trial wave function as 
\begin{equation}
    \Psi_T(\mathbf{X}) = e^{U(\mathbf{R})} \sum\limits_\alpha c_\alpha \textrm{det}_\alpha \left[ \ldots, \chi_i(\mathbf{r}_k,s_k) ,\ldots \right].
\end{equation}
with $i,k=1,...,N$.
The particle correlations are explicitly approximated by the Jastrow factor $ U(\mathbf{R})$ that captures two-particle and,
possibly, higher order correlations, as customary 
in QMC calculations \cite{qmcrev,acta,kolorenc,melton2}.

\subsection{From fixed-phase to fixed-nodes}

In this work we are particularly focused on the limit of vanishing spin-orbit and how 
the single-reference spinor determinant simplifies 
to the product of spin-up and spin-down determinants, i.e., to the usual fixed-node form. 
Let us show that this is indeed what happens for our spin representation as briefly sketched earlier \cite{melton2}.
Note that our previous exposition of this aspect was not formulated precisely \cite{melton2}, 
so that we clarify it in detail here. For the sake of consistency with the previous paper we
consider $N$ occupied spinors that can be grouped
as $N/2$ Kramer's pairs (for simplicity assuming $N$ to be even). 
 We can write the Kramer's pair as
 \begin{eqnarray}
\chi^+ &=&(\varphi+\Delta\varphi)\chi^{\uparrow}+
(\varphi-\Delta\varphi)\chi^{\downarrow} \\
\chi^- &=& (\varphi-\Delta\varphi)\chi^{\uparrow}-
(\varphi+\Delta\varphi)\chi^{\downarrow}
\end{eqnarray}
where the $\Delta\varphi$ is the spin-orbit induced splitting 
of the spatial orbital $\varphi$. The block of the first four rows/columns 
from the corresponding Slater determinant reads as follows
\begin{equation}
{\rm det} \left[\begin{matrix}
\chi_1^+(1) & \chi_1^+(2)& \chi_1^+(3)& \chi_1^+(4)&... \\
\chi_1^-(1) & \chi_1^-(2)& \chi_1^-(3)& \chi_1^-(4)&... \\
\chi_2^+(1)& \chi_2^+(2)& \chi_2^+(3)& \chi_2^+(4) &... \\
\chi_2^-(1) &  \chi_2^-(2)& \chi_2^-(3)& \chi_2^-(4)&... \\
 & ... &  & & \\
\end{matrix}\right].
\end{equation}
Now we assume that the spin-orbit splitting $\Delta\varphi \to 0$ and then 
\begin{eqnarray}
\chi^+&=& \varphi (e^{is}+e^{-is}) \to \varphi e^{is}\\
\chi^-&=& \varphi (e^{is}-e^{-is}) \to \varphi e^{-is}
\end{eqnarray}
by elementary rearrangements (adding, subtracting rows with the same $\varphi$). Explicitly, this gives
\begin{equation}
{\rm det} \left[
\begin{matrix}
\varphi_1(1)e^{is_1} & \varphi_1(2)e^{is_2} & \varphi_1(3)e^{is_3}& \varphi_1(4)e^{is_4}&
... \\
\varphi_1(1)e^{-is_1}& \varphi_1(2)e^{-is_2}& \varphi_1(3)e^{-is_3}& \varphi_1(4)e^{-is_4}& ...
\\
\varphi_2(1)e^{is_1} & \varphi_2(2)e^{is_2}& \varphi_2(3)e^{is_3}& \varphi_2(4)e^{is_4}& ... \\
\varphi_2(1)e^{-is_1}& \varphi_2(2)e^{-is_2}& \varphi_2(3)e^{-is_3}& \varphi_2(4)e^{-is_4}& ... \\
 & ... & & & \\
\end{matrix}
\right]
\end{equation}
This effectively complexifies the usual real wave function with the additional difference that the Slater
matrix is of size $N\times N$.
Clearly, for arbitrary spins variables this wave function is {\em different} from the usual 
spin-up and spin-down product although in what follows we will demonstrate how to recover such the fixed-node form using 
an appropriate choice for the spin coordinates.
Let us now assume that $s_i=s_1,s_3, ... $ will become the spin-up channel while $s_i=s_2,s_4, ... $ will end up being the spin-down channel. In order to reach this spin-up/down partitioning explicitly we restrict
$s_1,s_3, s_5, ...,=s$, and $s_2,s_4, ...=s' $ where $s,s'$ are distinct. Then we can write the determinant

\begin{equation}
{\rm det} \left[\begin{matrix}
\varphi_1(1)e^{is} & \varphi_1(2)e^{is'} & \varphi_1(3)e^{is} & \varphi_1(4)e^{is'} & ...   \\
\varphi_1(1)e^{-is}& \varphi_1(2)e^{-is'} & \varphi_1(3)e^{-is}& \varphi_1(4)e^{-is'}& ... \\
\varphi_2(1)e^{is}& \varphi_2(2)e^{is'}& \varphi_2(3)e^{is}& \varphi_2(4)e^{is'}& ... \\
\varphi_2(1)e^{-is}& \varphi_2(2)e^{-is'}& \varphi_2(3)e^{-is}& \varphi_2(4)e^{-is'}& ... \\
 & ... &  & & \\
\end{matrix}\right]
\end{equation}
and eliminating elements in each odd row
\begin{equation}
{\rm det} \left[\begin{matrix}
0 & c_0\varphi_1(2) & 0 &c_0\varphi_1(4)& ...   \\
\varphi_1(1)e^{-is} & \varphi_1(2)e^{-is'}& \varphi_1(3)e^{-is}& \varphi_1(4)e^{-is'}& ... \\
0 & c_0\varphi_2(2)& 0 & c_0\varphi_2(4)& ... \\
\varphi_2(1)e^{-is}& \varphi_2(2)e^{-is'} & \varphi_2(3)e^{-is}& \varphi_2(4)e^{-is'} & ... \\
&  ... &  &  & \\
\end{matrix}\right]
\end{equation}
where 
%$$c_0=[e^{is'}e^{-2is}-e^{-is'}]=e^{-is}[e^{i(s'-s)}-e^{-i(s'-s)}] \propto \sin(s-s').$$ 
$$c_0=[e^{is'}-e^{i(2s-s')}]=e^{is}[e^{i(s'-s)}-e^{-i(s'-s)}] = 2ie^{is}\sin(s'-s).$$ 
Furthermore, by reshuffling the first two rows and columns and factorizing out the spins 
from the determinant 
we get
\begin{equation}
\propto[\sin(s'-s)]^{N/2}{\rm det} \left[\begin{matrix}
\varphi_1(1) &\varphi_1(3)& 0 & 0 & ... \\
 \varphi_2(1)& \varphi_2(3)& 0 & 0 & ...\\
0 & 0 &\varphi_1(2) &\varphi_1(4)& ... \\
0& 0& \varphi_2(2)& \varphi_2(4)& ...\\
 & ... & & & \\
\end{matrix}\right].
\end{equation}
After reshuffling the rest of rows and columns, the single determinant of spinors factorizes into the product of two determinants of spin-up and spin-down
block matrices. Generalization to odd $N$ with unpaired spinor(s) is straightforward. Therefore this decomposition strictly depends on the fact that 
all the spins have to acquire one of the two distinct values as expected when going from continuous
to the fixed-label representation.

\subsection{Wave functions with full space-spin symmetries}

In our recent paper we have probed into the behavior of such wave functions for simple cases \cite{melton3}. 
It is useful to use an example such as the
Li atom wave function to illustrate various wave function forms we consider here (assuming usual nucleus-electrons Hamiltonian without spin terms).
The full symmetry exact wave function
for the Li atom doublet is given by\cite{white}
\begin{eqnarray}
\Psi(1,2,3) &=& |\uparrow\rangle_1 |\uparrow\rangle_2
 |\downarrow\rangle_3 F(1,2,3)
 \nonumber\\
 &+&|\uparrow\rangle_1 |\downarrow\rangle_2
 |\uparrow\rangle_3F(3,1,2) \nonumber\\
 &+&
 |\downarrow\rangle_1 |\uparrow\rangle_2
 |\uparrow\rangle_3F(2,3,1) 
\end{eqnarray}
where the function $F$ depends only on the spatial coordinates. The function $F$ is the exact, irreducible,
spatial variables-only eigenstate for the three electrons in the doublet state. Indeed, it corresponds 
to the exact fixed-node solution sought after, say, in the FNDMC method. 

The single-configuration trial wave function in 
the fixed-node framework would look like 
\begin{equation}
\Psi_T(1,2,3)=\textrm{det}^{\uparrow}_{1,2}[1s,2s]\textrm{det}^{\downarrow}_{3}[1s]
\end{equation}
where the electrons 1 and 2 are assigned as spin-up while the electron 3 is spin-down. Clearly, this is just 
a projection onto the spin state 
$|\uparrow\rangle_1 |\uparrow\rangle_2
 |\downarrow\rangle_3
$
with the single-reference term approximating the spatial
part $F(1,2,3)$.

Our wave function with variable spins is given by
\begin{eqnarray}
\Psi &=&  \textrm{det}[1s\times e^{is},1s\times e^{-is},2s\times e^{is}] \nonumber \\
&=& e^{i\Phi_1}\textrm{det}^{\uparrow}_{1,2}[1s,2s]
\textrm{det}^{\downarrow}_{3}[1s] - e^{i\Phi_2}
\textrm{det}^{\uparrow}_{3,1}[1s,2s]\textrm{det}^{\downarrow}_2[1s] \nonumber \\
&&+ e^{i\Phi_3}\textrm{det}^{\uparrow}_{2,3}[1s,2s]
\textrm{det}^{\downarrow}_1[1s].
\end{eqnarray}
It therefore results in determinantal approximations to the function $F$ with the phase factors from varying spins as coefficients.
If one chooses, $s_1 = s_2 = s$ and $s_3 = s'$, the wave function collapses to a single determinant with a spin variable dependent coefficient as described above.

Sampling both spin and position spaces enables one to evolve between the spatial wave functions with permuted
coordinates,
i.e., eventually sampling all such equivalent possibilities.  
Note that the overall structure of the exact wave function and our variable spin formulation are 
analogous. While in this example the variable spins 
and corresponding phase factors
appear superfluous, the form becomes 
fully meaningful whenever spin-dependent terms in the Hamiltonian are
switched on.

\section{Fixed-phase variable spins QMC as a general method}
In fixed-node QMC calculations with real wave functions
the node improvement is often very  
challenging since any general method proposed so far appears 
to have very unfavorable scaling. In several papers we have
made some progress in understanding the relations between
electron density, multiplicity of bonds and node curvatures
that appear to be related with increased fixed-node bias
\cite{rasch}.
In addition, we found relationships between nodes and eigenvalues 
 that show the nodes carry information 
about the spectrum as presented elsewhere  \cite{melton5}. 

In this respect, the fixed-phase approximation opens new perspectives both in a better understanding of related issues 
with regard to antisymmetry and the corresponding fermion sign problem as well as possibilities for new constructions 
of more efficient approximations. 

One important property of the fixed-phase approximation is that the sampled distribution $\rho$ is non-negative everywhere and, as we mentioned, generically its zero locus is 
 a subset of configurations 
with codimension 2, i.e., 
{\em two} dimensions lower than the full configuration space.
In that case the sampling is of the configuration space is ergodic. One then solves for the bosonic ground state in a given, state-dependent potential. A simple toy example is an
atomic two-particle $^3P$ state with the wave function 
$$
\Psi(1,2)=r_1r_2 g(r_1)g(r_2)[Y_{11}(1)-Y_{11}(2)] 
$$
where $g$ are positive radial functions. 
Its phase-generated potential is given by
$$
V_{ph}={1\over 2[(x_1-x_2)^2+(y_1-y_2)^2]}
$$
while the corresponding non-negative amplitude $\rho(r_1,r_2)$ vanishes only at $x_1=x_2, y_1=y_2$
\cite{melton3}. This has also other consequences that make it 
favorable in comparison with the fixed-node approach, namely,
the divergences of the local energy and drift are significantly
diminished making them much smoother. For example,
the drift for the importance sampled distribution given by $\nabla\ln\rho$ 
is smooth except at the point of vanishing $\rho$. This removes complications 
around nodes of real functions such as large local energy fluctuations, non-zero probability of 
crossing/re-crossing the node within a given time step, possible occurrences of stuck walkers and others, due to the 
fact that $\ln \Psi_T$ is non-analytic at the node. All these difficulties can be brought under the control by decreasing the time step in the fixed-node formalism. However, here these complications are simply absent in the fixed-phase formulation by being smoothed out into the lower dimension.

We note that in low-dimensional systems or for particular symmetry constraints one can end up with 
special or non-generic cases having zeros of $\rho$ with codimension 1, i.e., one dimension higher than the generic codimension 2 mentioned above. A simple example is the lowest two-particle triplet in a periodic box with the wave function
${\rm det}[1,e^{ikx}]$. This leads to 
$\rho(x_1,x_2)= 2|\sin[(x_1-x_2)/2]|$ that has a $(2d-1)$-dimensional zero locus regardless of $d$, i.e., the dimensionality of the box.
The reason is that this particular state effectively behaves as having 1D nodal structure that is non-generic.
Interestingly enough, for $d>1$ this node volume is smaller than in the corresponding fixed-node wave function given by the real (or imaginary) part, ${\rm Re}\{{\rm det}[1,e^{ikx}]\}$. 
This aspect is more thoroughly 
investigated in our subsequent work
\cite{melton5} that explore the corresponding properties 
of nodes in such cases and further generalizations. 

Perhaps the most appealing and yet unexplored property is that the approximation has
a form of an additive effective many-body potential
\begin{equation}
V_{ph}=(1/2)[\nabla\Phi_T]^2
\end{equation}
so that the original Schr\"odinger equation changes to
\begin{equation}
(T+V)\Psi=E\Psi \quad \to \quad (T+V+V_{ph})\rho=E\rho
\end{equation}
This effective potential formulation  offers
a clear conceptual understanding of the transformed problem
that reminds us of effective potential/field methods used in other
areas of quantum and high energy physics. It has a number of 
desired properties when thinking about the solution of
the many-body problem, such
as that the solution is non-negative everywhere, the state-dependent potential $V_{ph}$ is purely repulsive (it only raises the energy) and it is explicitly and directly given by the approximate phase.
Consequently, it provides a constructive path for 
improvements with the perspective that the solution 
really exists, i.e., in the case of the exact 
phase one obtains the exact solution/eigenstate similarly to the fixed-node approximation (that is its special case). 
Interestingly, not much is known about 
the phases of stationary states.
It is possible that more thorough
analysis of the corresponding
effective potentials will lead to a better
understanding as well as to better approximations
for practical calculations of realistic systems.
What follows provides the first attempts to probe
some aspects of this formulation.

\section{Results}
We calculate the total energies for the first-row atoms using both the FN and FP approximations.
For the FN calculations, we build our trial wave function from HF orbitals generated from \textsc{Gamess-US} \cite{GAMESS}.
For the FP calculations, we build our trial wave function from the one-particle spinors generated from \textsc{Dirac14} \cite{DIRAC}.
For the FN calculations, we perform a linear time step extrapolation to zero time step. 
In all cases, a spatial time step of 0.001~Ha is in agreement with the zero time step limit.
Motivated by that, for all FP calculations we hold the spatial time step fixed at $\tau = 0.001$~Ha, rather than performing a spatial time step extrapolation.

We have previously studied some of the aspects of spin time steps \cite{melton,melton2}. 
%{\bf Rewrite/adjust the following: }
The analysis of total energies as function of spin time step leads to the following conclusions:

a) At large spin time step the spins are basically fully integrated out for each spatial step that is assumed to be much smaller. 
Then one sees higher fixed-phase bias since the repulsive potential acts in the full configuration space unlike the fixed-node condition that applies only on the configuration subspace.  

b) At very small spin steps and for small number of electrons, the propagation eventually finds the region(s)
close to the pure fixed-node wave function. Apart from small spin fluctuations the energy therefore reaches
very closely to the fixed-node solution.

In both limits and also for intermediate time step
regimes the energy is an upper bound, since the energy basically limited 
from below by the fixed-node limit. 
The complexified wave function and the fixed-phase only increases the energy
since it acts in full space instead of fixed-node codimension 1 hypersurface and expands the configuration 
space in an ad hoc manner through the continuous spin
as we argued in previous parts.  
%The argument here is based on the fact that by 
%expanding the configuration space the mixing of excitations in corresponding trial function will only increase.
This has also further implications that single reference wave functions will be, in general, less accurate, as we have 
actually observed in calculations of several systems \cite{melton2,melton3}.  
Here we are actually focused on the short spin time-step 
limit that enables us to recover the fixed-node results
although the calculations are carried out in FPSODMC
setting.
%{\bf up to here}

As described in \S \ref{section:wf}, we initialize the spin-configurations to facilitate the decomposition into two independent determinants, as must be the case in a spin independent Hamiltonian. 
As an illustration of why this is necessary, consider the N atom using HF spatial orbitals and no Jastrow factor. 
The VMC energy should agree with the HF energy, within the statistical errors. 
If one randomly generates the spin variables and performs a VMC calculation, the obtained energy is $-54.3341(8)$~Ha which clearly disagrees with the HF value of -54.40093 Ha. 
However, if we initialize the spin variables such that $s_1 = s_3 = s_5 = s_6 = s_7 = s$ and $s_2 = s_4 = s'$ while using a very small spin time step $\tau_s$, we obtain $-54.4003(7)$~Ha, which agrees with the energy obtained via HF. 
By initializing the spin variables such that the wave function properly decomposes into a product of determinants, we can use a small $\tau_s$ such that the spin variables stay close to the original configurations.
We allow the spin variables to continue to drift in order to sample the spin configuration space.

\begin{figure}[h]
    \centering
    \includegraphics[width=0.5\textwidth]{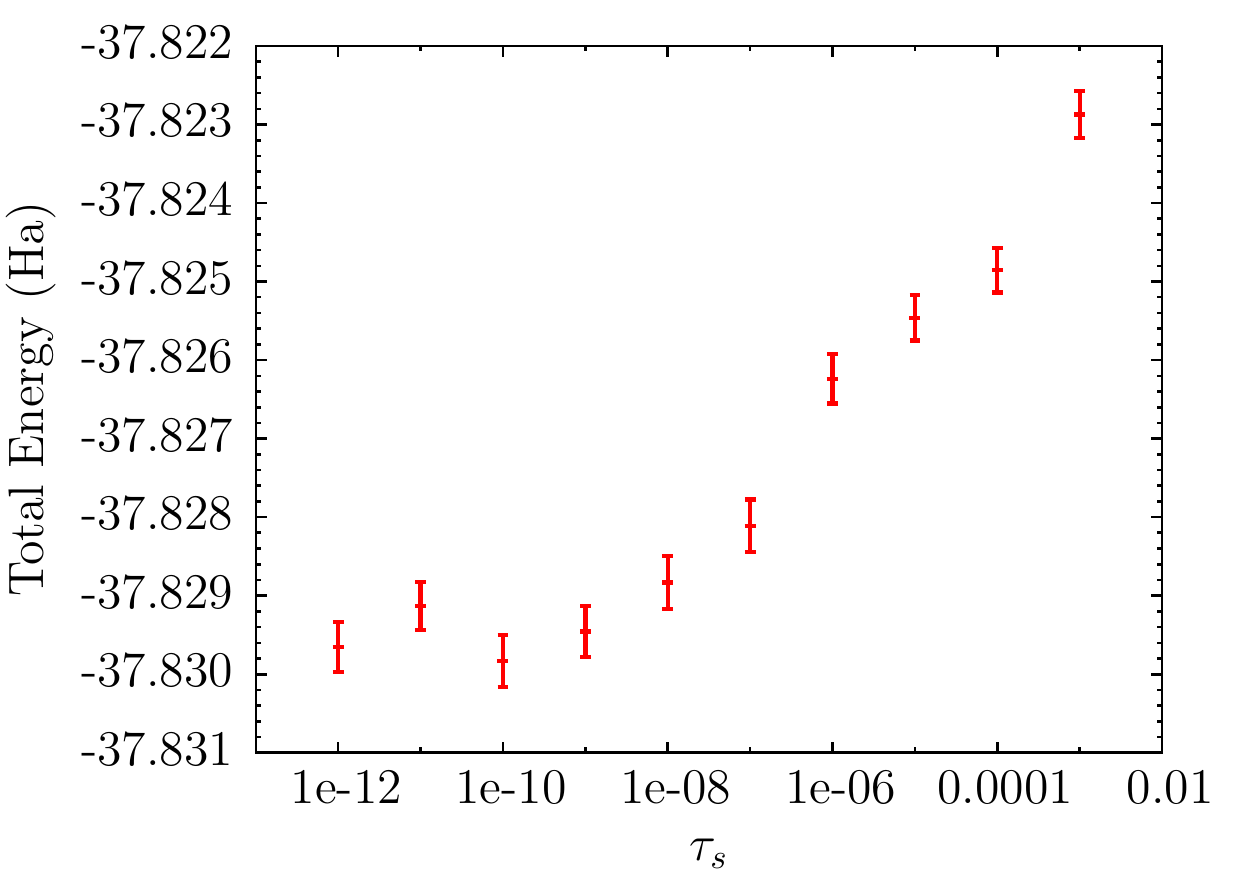}
    \caption{FPDMC energy of the C atom with varying spin time steps $\tau_s$. The initial spin configurations were chosen to in order to decompose into a product of determinants. }
    \label{fig:convergence}
\end{figure}

\begin{figure}[h]
    \centering
    \includegraphics[width=0.5\textwidth]{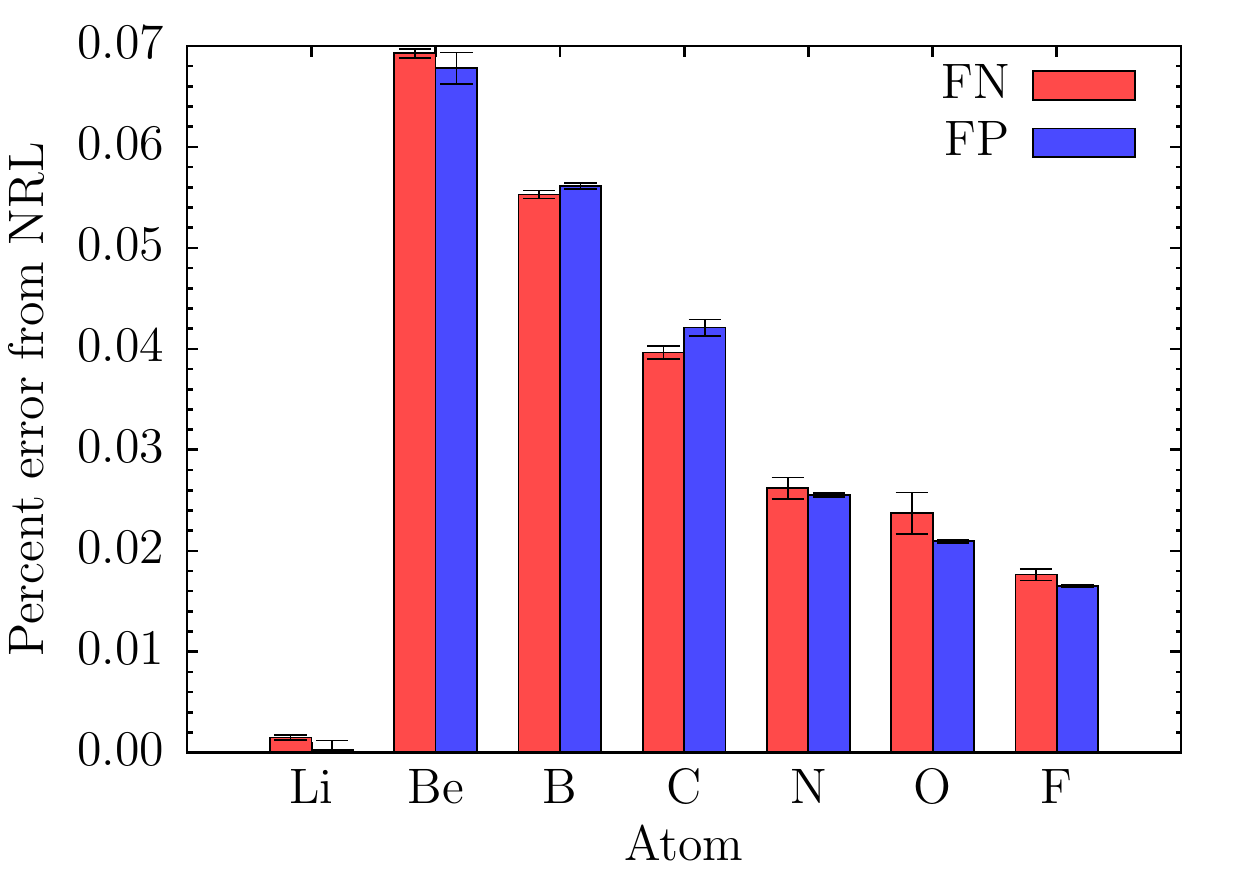}
    \caption{Percentage error on the total energies in the fixed-node (FN) and fixed-phase (FP) extrapolated calculations.}
    \label{fig:errors}
\end{figure}

When performing FPSODMC, we vary the spin time step until the energy is saturated for a fixed spatial time step. 
An example of the spin time step extrapolation is shown in Figure \ref{fig:convergence}.
For $\tau_s$ between $10^{-12}$ and $10^{-9}$, the DMC energies all agree to within the error bars. 
Performing the same procedure for all atoms, we list the total energies in Table \ref{tab:total_energy}.
By comparing the FN and FP total energies to the ``exact'' energies in the non-relativistic limit (NRL)\cite{nrl}, we calculate the fixed-node/phase error as the percentage of the total as it is plotted in Figure \ref{fig:errors}.

Regardless of the approximation, the associated error decreases with atomic number subject to the choice of HF nodes/phases. 
Additionally, the FN and FP approximations yield essentially identical errors.

\begin{table*}
    \centering
    \begin{tabular}{crrrr}
	\hline\hline
	Atom  &    HF     &   NRL\cite{nrl}    &     FN   &   FP   \\ 
	\hline
	Li & $-7.43272$ & $-7.47806$     & $ -7.47794(2)$      &  $    -7.47804(7)   $  \\    
	Be & $-14.57302$ & $ -14.66736 $ & $  -14.65720(6) $   &  $   -14.6574(2)    $  \\
	B  & $-24.52906$ & $ -24.65393 $ & $  -24.64030(9) $   &  $   -24.64016(8)   $  \\
	C  & $ -37.68861$ & $ -37.84500 $ & $  -37.8300(3) $   &  $   -37.8291(3)    $  \\
	N  & $ -54.40093$ & $ -54.58930 $ & $  -54.5750(6) $   &  $   -54.5754(1)    $  \\
	O  & $ -74.80939$ & $ -75.06700 $ & $  -75.049(1)  $   &  $   -75.0513(1)    $  \\
	F  & $ -99.40934$ & $ -99.73400 $ & $  -99.7164(6) $   &  $   -99.7175(1)    $  \\
	\hline
    \end{tabular}
    \caption{Total energies in Ha for the first-row elements using FN(FP) DMC with HF nodes (phases). FN calculations are extrapolated to zero time step. FP calculations take a spatial time step of 0.001 and decrease the spin time step until the energy is unchanged. NRL is the estimated nonrelativistic exact energy.}
    \label{tab:total_energy}
\end{table*}

\begin{figure}[h]
    \centering
    \includegraphics[width=0.5\textwidth]{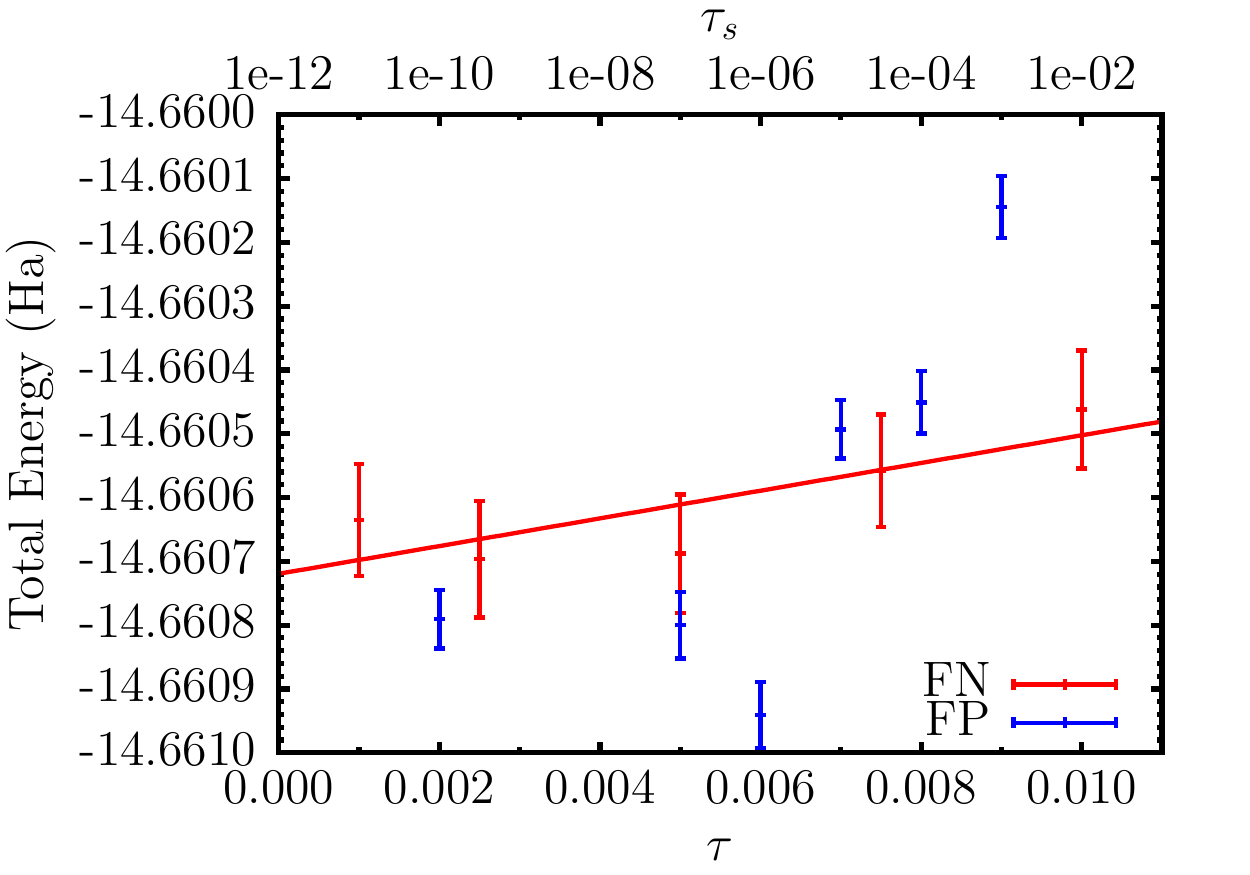}
    \caption{Total energies for the Be atom with a two-configuration wave function. The bottom $x$-axis shows the real spin time step, which is linearly extrapolated to zero time step with an energy of -14.66071(5) Ha. The top $x$-axis indicates the spin time step, where with each value configurations are initialized such that $s_1 = s_3 = s$ and $s_2 = s_4 = s'$. }
    \label{fig:be_csf}
\end{figure}

It is well known that using HF nodes, Be has a significant FN error \cite{rasch}.
The ground state symmetry of Be is $^1S_0$, which is obtained with the electron configuration $1s^2 2s^2$. 
Using a HF trial wave function, the nodal surface $\partial \Omega = \left\{ \mathbf{R} \in \mathbb{R}^{3N} | \Psi_T(\mathbf{R}) = 0 \right\}$ separates the configuration space into 4 nodal domains, two of which where the wave function is positive and two in which it is negative. 
It is well-known that by adding just one more configuration that is related to the near-degenerate state
of the same symmetry one finds only two nodal domains 
as expected for generic fermionic ground state
\cite{ceperley1991,bressanini2002,mitas2006}. Previous calculations have found that it almost completely eliminates 
the fixed-node bias \cite{umrigar1988, bressanini2002}. The corresponding two-configuration trial function is given by 
\begin{equation}
    |\Psi_T\rangle = c_0 |1s^2 2s^2\rangle + c_1 \sum\limits_{i \in \{x,y,z\}} | 1s^2 2p_i^2\rangle
    \label{eqn:multi-Be}
\end{equation}
as 
With this choice of trial wave function
and full optimization all variational parameters one can reach FN result with almost zero bias \cite{umrigar1988}.
Instead we perform an optimization of this wave function with only the Jastrow parameters and expansion coefficient, keeping the HF orbitals fixed with resulting small increase in the energy compared to the nearly exact value.
Total energies are shown in Figure \ref{fig:be_csf}.
Again, by choosing the FP calculations to preserve the spin assignments of $s_1 = s_3 = s$ and $s_2 = s_4 = s'$ by using a small spin time step, the FN and FP calculations agree to within statistical uncertainty.

Thus far, we have only presented results for all-electron systems.
We also consider the FN and FP approximation when nonlocal pseudopotentials are included \cite{melton,melton2}. 
We calculate the binding curve for the nitrogen dimer in the $^1 \Sigma_g$ state, using a single-reference trial wave function for each approximation built from HF spatial orbitals.
We utilize a BFD pseudopotential for N \cite{bfd}. 
In order to calculate the binding curve, we first calculate the isolated N atom both in FN and FP.
Under the locality approximation \cite{mitas91}, we perform a time step extrapolation within FNDMC and obtain a total energy of $-9.7912(1)$~Ha.
Using a fixed spatial time step, we perform a spin time step extrapolation as described above to facilitate decomposition into a product of two independent determinants using FPSODMC and obtained a total energy of $-9.7917(4)$~Ha.

\begin{figure}[h]
    \centering
    \includegraphics[width=0.5\textwidth]{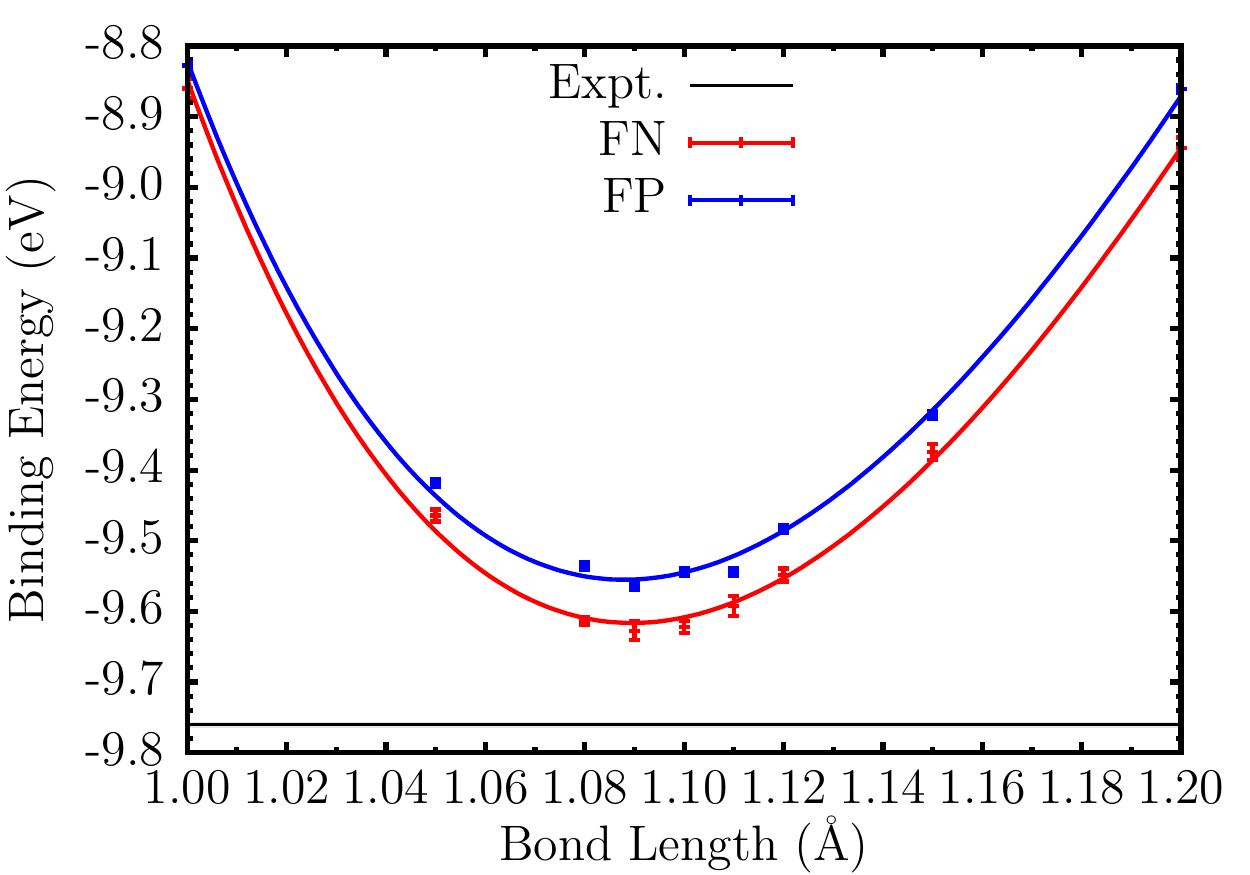}
    \caption{N$_2$ binding curve for the $^1\Sigma_g$ molecular state using a HF nodal surface/phase. The horizontal line indicates the experimental dissociation energy. The experimental error bar is too small to be visible on this scale.
The small increase in FP underbinding comes from a slightly smaller fixed-phase bias in the N atom and a slightly larger bias in the dimer with the HF phase. The PBE0 phase calculations lower the binding curve minimum further and make the difference with the
FN results even smaller, see text. 
   }
    \label{fig:dimer}
\end{figure}

The dimer curve is shown in Figure \ref{fig:dimer}
and shows the binding obtained from the FN and FP methods.
The QMC data is fit to the morse potential
\begin{equation}
V(r) = D_e\left[ e^{-2a(r-r_e)} -2e^{-a(r-r_e)} \right]
    \label{eqn:morse}
\end{equation}
and the vibrational frequency can be obtained via
\begin{equation}
    \nu_0 = \frac{1}{2\pi}\sqrt{\frac{2a^2D_e}{\mu}}
    \label{eqn:frequency}
\end{equation}
where $\mu$ is the reduced mass of the dimer.
The FP solution for the dimer has a slightly larger bias in comparision to the FN solution,
on the order of 1~mHa across the entire binding curve. 
Coupled with the {\em slightly} lower energy for the individual atom, the overall binding energy differs from the FN result by roughly $\sim$~0.1~eV, as shown in Table \ref{tab:dimer}.
For completeness, we calculated the dimer using an improved nodal surface/phase given by a trial wave function composed of PBE0 nodes at the equilibirum bond length. 
At $r_e = 1.09$~\r{A}, the PBE0 nodal surface is lower in energy by only 0.0010(5)~Ha, which slightly improves the binding energy prediction to $9.654(8)$~eV.
The PBE0 phase has a more significant improvement over the HF phase, yielding a lower energy by 0.0029(5)~Ha. 
The binding energy prediction becomes $9.64(1)$~eV, which is very close to the FN result.
Clearly, the differences between the methods are very small,
basically similar to variations in the fixed-node 
biases for different atoms and molecular systems and choices of orbitals used in single-reference trial functions. 

\begin{table}
    \centering
    \caption{Equilibrium bond lengths ($r_e$), dissociation energies ($D_e$) and vibrational frequencies ($\nu_0$) for the various approximations compared to experiment using a HF nodal surface/phase. Parameters and uncertainties are obtained from a fit to the Morse potential.}
    \begin{tabular}{c c c c }
	\hline\hline
	Method & $r_e$ (\r{A}) & $D_e$ (eV) & $\nu_0$ (cm$^{-1}$) \\
	\hline
	FN & 1.0895(8) & 9.616(5) & 2402(28)\\
        FP & 1.0879(7) & 9.555(6) & 2396(23) \\
	Expt.\cite{nist} & 1.098 & 9.758(6) & 2358.57(9) \\
	\hline
    \end{tabular}
    \label{tab:dimer}
\end{table}

\section{Conclusions}\label{section:conclusions}

In this paper we elaborate in detail on a particularly important aspect of the fixed-phase spin-orbit/spinors DMC (FPSODMC) method that we have introduced recently \cite{melton}.
We highlight some of the key aspects, in particular, how to obtain the fixed-node limit results from the fixed-phase setting both in theory and in practical calculations. 
We point out the promising features of the fixed-phase 
method and also show its behavior in our continuous spin formalism. We illustrate the results on first row atoms calculations.
  The method enables us to write 
full space-spin symmetry wave functions for Hamiltonians 
with or without explicit spin terms and opens thus
possibilities for further improvements of trial wave functions.
We consider the results very encouraging since 
in a straightforward manner we were able to obtain the 
fixed-node results in both all-electron and effective core potential
settings as well as confirm essentially the same quality of both single and multi-reference trial wave functions.
The method opens interesting new perspectives for 
many-body electronic structure calculations in complex wave function and spinor formalism
that takes into account variable nature of the spin degrees of freedom and provides new possibilities for construction
of more general trial wave 
functions. 

{\em Acknowledgments.}
This research was supported by the U.S. Department of Energy (DOE), Office of Science, Basic Energy Sciences (BES) under Award DE-SC0012314. For calculations we used resources at NERSC, a DOE Office of Science User Facility supported by the Office of Science of the U.S. Department of Energy under Contract No. DE-AC02-05CH11231.
 Most calculations have been carried out at TACC.

\bibliography{bibfile}
\end{document}